\documentclass[12pt]{iopart}

\usepackage{amsmath}
\usepackage{amsfonts, amssymb}
\usepackage{booktabs}
\usepackage{graphicx}
\usepackage{subfigure}
\usepackage{soul, color}
\begin{document}
\title[]{Theoretical study of Magnetoresistance Oscillations in Semi-parabolic Plus Semi-inverse Squared Quantum Wells in the Presence of Intense Electromagnetic Waves}
\author{Nguyen Thu Huong$^1$,Nguyen Quang Bau$^{2,*}$, Cao Thi Vi Ba$^{2}$, Bui Thi Dung$^{2}$, Nguyen Cong Toan$^{2}$, and Anh-Tuan Tran$^{2, \dagger}$}
\address{$^1$Faculty of Basic Science, Air Defence-Air Force Academy, Kim Son, Son Tay, Hanoi, Vietnam}
\address{$^{2}$Department of Theoretical Physics, University of Science, Vietnam National University, Hanoi, Address: $\rm{No}$ 334 Nguyen Trai, Thanh Xuan, Hanoi, Vietnam}
\ead{$^{*}$nguyenquangbau54@gmail.com, nguyenquangbau@hus.edu.vn}
\ead{$^{\dagger}$trananhtuan\_sdh21@hus.edu.vn}
\vspace{5pt}
\begin{indented}
\item[]November 2024
\end{indented}

\begin{abstract}
Magnetoresistance oscillations in semiconductor quantum wells, with the semi-parabolic plus semi-inverse squared potential, under the influence of intense electromagnetic waves (IEMW), is studied theoretically. Analytical expression for the longitudinal magnetoresistance (LMR) is derived from the quantum kinetic equation for electrons, using the Fröhlich Hamiltonian of the electron-acoustic phonon system. Numerical calculation results show the complex dependence of LMR on the parameters of the external field (electric, magnetic field and temperature) as well as the structure parameters of the confinement potential. In the absence of IMEW, Shubnikov-de Haas (SdH) oscillations appear with amplitudes that decrease with temperature in agreement with previous theoretical and experimental results. In the presence of IEMW, the SdH oscillations appear in beats with amplitudes that increase with the intensity of the IEMW. SdH oscillations under the influence of electromagnetic waves are called microwave-induced magnetoresistance oscillations. The maximum and minimum peaks appear at the positions where the IEMW frequencies are integer and half-integer values of the cyclotron frequency, respectively. In addition, the structural parameters of the quantum well such as the confinement frequency and the geometrical parameters have a significant influence on the LMR as well as the SdH oscillations. When the confinement frequency is small, the two-dimensional electronic system in the quantum well behaves as a bulk semiconductor, resulting in the absence of SdH oscillations. In addition, the LMR increases with the geometrical parameter $\beta_z$ of the quantum well. The obtained results provide a solid theoretical foundation for the possibility of controlling SdH oscillations by IEMW as well as the structural properties of materials in future experimental observations.
\end{abstract}

\section{Introduction}
In recent years, the magneto-optical properties of low-dimensional electronic systems have increasingly attracted widespread attention from many researchers, both theoretically \cite{lt1,lt2,lt3} and experimentally \cite{tn1}. Among them, two-dimensional electronic gases such as quantum wells, superlattices and Dirac materials showing unique electronic properties that have the potential for a wide range of applications in new miniaturized nano-electronic components \cite{ap1,ap2,ap3}. It is well-known that two-dimensional electronic gases are a scientific model in which the electron gas is strongly confined in one dimension (assuming the Oz axis, called the growth direction) and moves freely in the remaining two dimensions. As a result, with different confinement potentials, electrical and optical properties of two-dimensional electronic gases are very different in comparison with normal bulk semiconductors \cite{in1, in5}. In particular, under the influence of an external magnetic field placed along the growth direction of two-dimensional electronic gases in free-standing semiconductor quantum well structures, the electron state is quantized into Landau levels \cite{in2, in3}. Indeed, the presence of a magnetic field is equivalent to the appearance of a parabolic one-dimensional potential energy, leading to the solution of the Schrodinger equation as a product of a plane wave function and a harmonic oscillation function at the center of the Landau orbit ${{x}_{0}}=-\ell _{B}^{2}{{k}_{y}}$, with $\ell_B$ being the magnetic length \cite{van1}. Many studies have been carried out to solve the problems of magneto-optical absorption in quantum wells with different profiles such as parabolic potential \cite{lt2}, infinite asymmetric potential \cite{fqw1, hsr1}, Pöschl-Teller potential \cite{in4}, and so on. However, the above studies have not yet considered the presence of a uniform transverse electric field and detailed discussions on the influence of magnetic and electric fields on magneto-electric effects are still left open. 

One of the most important magneto-electric effects in two-dimensional electronic gases under the influence of a strong magnetic field is the quantum Hall effect discovered by Klaus V. Klitzing in 1980 \cite{qhe1}. The quantum Hall effect sets a new standard that allows applications in the metrology of important kinetic quantities such as the quantized Hall resistance, momentum relaxation time, effective mass, mobility of carriers, and so on \cite{qhe2}. In the low temperatures, quantum oscillations of longitudinal magnetoresistance (LMR) can be observed at the lower magnetic fields, called Shubnikov–de Haas (SdH) oscillations. E. Tiras \textit{et al.} \cite{qhe2} used SdH effect measurements to study the energy relaxation of electrons in $\text{A}{{\text{l}}_{\text{0}\text{.25}}}\text{G}{{\text{a}}_{\text{0}\text{.75}}}\text{N/AlN/GaN}$ heterostructures showing a monotonic decrease of the energy relaxation time with temperature in the acoustic phonon  regime at low temperatures via piezoelectric scatterings. In addition, the authors also experimentally investigated the temperature dependence of LMR and compared it with theoretical predictions of N. Balkan \textit{et al.} \cite{linke, ampli2} in $\text{GaAs}/\text{A}{{\text{l}}_{x}}\text{G}{{\text{a}}_{1-x}}\text{As}$ multiple quantum wells. The research results indicated that the relative amplitude reduces as the temperature rises, with respect to a constant magnetic field. Inspired by this, similar results were found in several recent theoretical studies in parabolic \cite{mr1} and rectangular \cite{mr3} quantum wells, and doped superlattices \cite{mr4}. Furthermore, to the best of our knowledge, there has not been any theoretical research on the magnetoresistance oscillations in quantum wells with asymmetric semi-parabolic potentials. Therefore, in this study, we choose to investigate SdH in asymmetric semi-parabolic quantum wells under the influence of intense electromagnetic waves (IEMW) to supplement the above shortcomings.

The purpose of choosing the semi-parabolic plus semi-inverse squared quantum well for study in this paper is based on two main reasons. First, the solution of the time-independent Schrödinger in semi-parabolic plus semi-inverse squared quantum well can be obtained in analytical form. This allows us to perform the analytical calculations explicitly. Second, advances in nano-fabrication techniques \cite{mbe1, mbe2, mbe3}, such as molecular beam epitaxy and metal-organic chemical vapor deposition, now enable the preparation of semiconductor quantum wells in various shapes, such as square \cite{fas1, fas2}, parabolic \cite{fap1, fap2, fap3}, triangular \cite{fatri1, fatri2, fatri3, fatri4} wells. These structures significantly influence electrical, optical, and transport properties. It is also expected that semi-parabolic plus semi-inverse squared quantum wells can be fabricated from heterostructures such as AlGaAs/GaAs using these advanced techniques. In recent years, previous theoretical studies \cite{hsr1, hsr2, hsr3, fqw1, fqw1p, fqw1pp} have been carried out and have brought many interesting new results on magneto-optical absorption effects and the dependence of absorption linewidth on the geometrical parameters of the semi-parabolic and semi-inverse squared quantum wells under the influence of IEMW. Inspired by these works, we continue to study the influence of structural parameters and strong electromagnetic waves on the Hall effect and magnetoresistance oscillations. We expect that our results will be considered useful predictions for future experiments as well as quantum well fabrication technology.

The influence of IEMW on the quantum Hall effect as well as SdH has been mentioned in many previous theoretical \cite{in6, qhemw, qhemw2, qhemw3} and experimental \cite{in7, in8} studies. The above studies indicate the appearance of microwave-induced magnetoresistance oscillations where the position of the maximum peaks of LMR is governed by the ratio of the cyclotron and the IEMW frequency \cite{mr1, mr3, mr4, qhemw, qhemw2}. In 1976, the research group of E.M. Epshtein and G.M. Shmelev \cite{ep1, ep2, ep3, ep4, ep5} first formulated the quantum kinetic equation \cite{tem} for the electrons in bulk semiconductors under the impact of combined the magnetic field and IEMW, establishing the basis for theoretical studies of the general quantum Hall effect and microwave-induced magnetoresistance oscillations in particular. The authors explained the appearance of microwave-induced magnetoresistance oscillations due to the nonlinear influence of IEMW on electric conductivity and electron-phonon scattering probability under low temperature and strong magnetic field conditions. These are the quantum limit conditions, where the classical Boltzmann kinetic equation is no longer valid. Therefore, on the basis of previous theoretical studies using the quantum kinetic equation \cite{mr1, mr2, mr3, mr4}, we continue to carry out calculations to clarify the influence of the external field, as well as the characteristic structural parameters of asymmetric semi-parabolic quantum wells on microwave-induced magnetoresistance oscillations. We present in detail and systematically the theoretical framework in the next section (Sec. 2). Then, in Sec. 3, we estimate numerical values, including an analysis of the dependence of LMR on the external fields, temperature and geometric structural parameters of quantum wells. Finally, conclusions are given in Sec. 4. 

\section{Theoretical Framework}
\subsection{Infinite semi-parabolic plus semi-inverse squared quantum wells and the time-independent Schrödinger equation for electrons}
In the effective mass approximation, the one-electron Hamiltonian under the perpendicular magnetic field $\mathbf{B}=\left( 0,0,B \right)$ and crossed electric field $\mathbf{E}_{c}=\left( E_x,0,0 \right)$ can be expressed as follows
\begin{align}\label{hamile}
    {{\mathcal{H}}_{e}}=\dfrac{{{\hbar }^{2}}}{2{{m}_{e}}}{{\left( \mathbf{k}+\dfrac{e}{\hbar }\mathbf{A} \right)}^{2}}+\mathcal{U}\left( z \right)+eE_{x}x,
\end{align}
with $e$, and $m_e$ are the charge and effective mass of electrons in Gallium Arsenide semiconductor (chemical formula GaAs). In the Landau gauge, the canonical momentum has the form $\boldsymbol{\Pi}=\hbar \left( \mathbf{k}+{e\mathbf{A}}/{\hbar }\; \right)$, in which, the magnetic vector potential $\mathbf{A}=\left( 0,Bx,0 \right)$. $\mathcal{U}\left( z \right)$ is the confinement potential of the infinite semi-parabolic plus semi-inverse squared quantum wells, its functional form is given by \cite{hsr1, hsr2, hsr3} $\mathcal{U}\left( z \right)=\dfrac{1}{2}{{m}_{e}}\omega _{z}^{2}{{z}^{2}}+\dfrac{{{\hbar }^{2}}{{\beta }_{z}}}{2{{m}_{e}}{{z}^{2}}},$  with  $\omega_z$ is the confinement frequency, characterizing the semi-parabolic part, while $\beta_z$ is the geometric structural parameter characterizing the inverse-square part of QWs. 

The time-independent Schrödinger equation for electrons in the infinite semi-parabolic plus semi-inverse squared quantum wells is given by 
\begin{align}
    {{\mathcal{H}}_{e}}{{\Psi }_{\text{N},\text{n},{{\mathbf{k}}_{y}}}}\left( \mathbf{r} \right)={{\mathcal{E}}_{\text{N},\text{n},{{\mathbf{k}}_{y}}}}{{\Psi }_{\text{N},\text{n},{{\mathbf{k}}_{y}}}}\left( \mathbf{r} \right).
\end{align}

After some manipulations, the total wave-function and corresponding
energy levels can be obtained as  \cite{hsr1, hsr2, hsr3, van2} 
\begin{align}
  & {{\Psi }_{\text{N},\text{n},{{\mathbf{k}}_{y}}}}\left( \mathbf{r} \right)=\dfrac{{{\text{e}}^{i{{k}_{y}}y}}}{\sqrt{{{L}_{y}}}}{{\Phi }_{\text{N}}}\left( x-{{x}_{0}} \right){{\phi }_{\text{n}}}\left( z \right), \\ 
 & {{\Phi }_{\text{N}}}\left( x-{{x}_{0}} \right)=\dfrac{{{\text{e}}^{-{{{\left( x-{{x}_{0}} \right)}^{2}}}/{2\ell _{B}^{2}}\;}}}{\sqrt{{{2}^{\text{N}}}\text{N}!\sqrt{\pi }{{\ell }_{B}}}}{{\text{H}}_{\text{N}}}\left( \dfrac{x-{{x}_{0}}}{{{\ell }_{B}}} \right), \\ 
 & {{\phi }_{\text{n}}}\left( z \right)=\sqrt{\dfrac{2\text{n}!}{\ell _{z}^{1+4\text{s}}\Gamma \left( 2\text{s}+\text{n}+\dfrac{1}{2} \right)}}{{z}^{2\text{s}}}{{\text{e}}^{-{{{z}^{2}}}/{2\ell _{z}^{2}}\;}}\mathcal{L}_{\text{n}}^{2\text{s}-{1}/{2}\;}\left( \dfrac{{{z}^{2}}}{\ell _{z}^{2}} \right), \\ 
 & {{\mathcal{E}}_{\text{N},\text{n},{{\mathbf{k}}_{y}}}}={{\mathcal{E}}_{\text{N},\text{n}}}-\hbar {{\upsilon }_{d}}{{k}_{y}}, \\ 
 & {{\mathcal{E}}_{\text{N},\text{n}}}=\left( \text{N}+\dfrac{1}{2} \right)\hbar {{\omega }_{B}}+\left( 2\text{n}+1+\dfrac{\sqrt{1+4{{\beta }_{z}}}}{2} \right)\hbar {{\omega }_{z}}+\dfrac{1}{2}{{m}_{e}}\upsilon _{d}^{2}, \label{enn}
\end{align}
here, ${{\rm{s}} = \dfrac{1}{4}\left( {1 + \sqrt {1 + 4{\beta _z}} } \right)}$, $L_y$ is the normalized length in the y-direction. ${{\Phi }_{\text{N}}}\left( x-{{x}_{0}} \right)$
represents the harmonic-oscillator wave functions, with the electric field $\bf{E}_{c}$ entering through the centre-coordinate
${{x}_{0}}=-\ell _{B}^{2}\left( {{k}_{y}}-{{{m}_{e}}{{\upsilon }_{d}}}/{\hbar }\; \right)$, in which, ${{\upsilon }_{d}}={E_x}/{B}\;$ is the drift velocity, and
${{\ell }_{B}}=\sqrt{{\hbar }/{\left( {{m}_{e}}{{\omega }_{B}} \right)}\;}$ is the effective magnetic length, with the cyclotron frequency ${{\omega }_{B}}={eB}/{{{m}_{e}}}\;$, ${{\ell }_{z}}=\sqrt{{\hbar }/{\left( {{m}_{e}}{{\omega }_{z}} \right)}\;}$. $\mathrm{H}_{\mathrm{n}}\left(x\right)$, $\Gamma\left( x \right)$ and $\mathcal{L}_{\text{n}}^{k}\left( x \right)$ are the n-th Hermite polynomials, Gamma functions and associated Laguerre polynomials. N is the Landau level index, and n is the electric subband index. These Landau levels are formed due to the quantization of the electron's motion perpendicular to the applied magnetic field. 
\subsection{Quantum kinetic equation for the electron distribution function} 
\begin{figure}[!htb]
    \centering
    \includegraphics[scale=0.8]{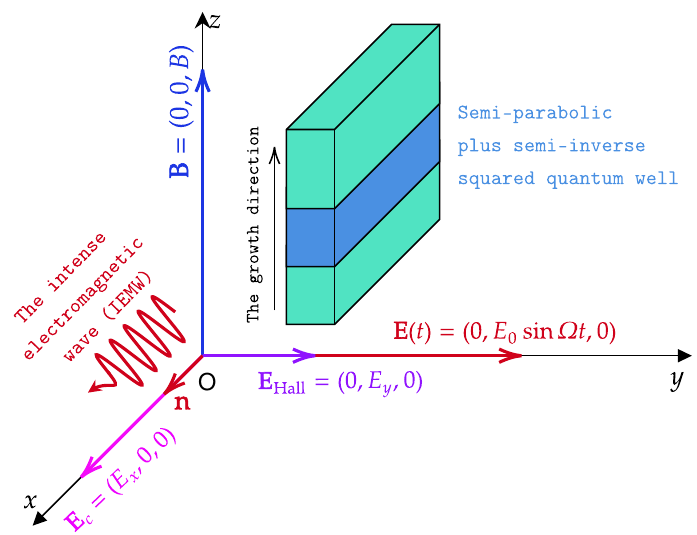}
    \caption{The diagram specifically illustrates the spatial orientation of the external fields such as electromagnetic waves, constant electric and magnetic fields.}
    \label{fig1}
\end{figure}

Besides external fields such as uniform electric and magnetic fields that affect the states and eigenvalues of the electron, we consider the infinite semi-parabolic plus semi-inverse squared quantum wells under the influence of intense electromagnetic waves propagating along the x axis, with the electric field vector ${{\mathbf{E}}\left( t\right)}=\left( 0,{{E}_{0}}\sin \Omega t,0 \right)$ (${{E}_{0}}$ and $\Omega$ are the amplitude and the frequency, respectively). Fig. \ref{fig1} is a simple schematic illustration of the quantum well model under the influence of external fields such as electromagnetic waves, constant electric and magnetic fields, in which we focus on clarifying the orientation of the external fields.

The quantum Hall effect occurs when a two-dimensional electronic system is confined in a potential well placed in a perpendicular magnetic field at low temperature. Under low temperature conditions, the electron-acoustic phonon scattering mechanism is dominant. Therefore, to consider in detail the effects of external fields when studying the quantum Hall effect, we start with the Fröhlich Hamiltonian of the electron-acoustic phonon system, which reads in second quantization formalism \cite{mr1, mr2, qke2, van1}
\begin{align}
   \mathcal{H}&={{\mathcal{H}}_{0}}+{{\mathcal{H}}_{\operatorname{int}}}, \label{ha1}\\ 
  {{\mathcal{H}}_{0}}&=\sum\limits_{\text{N},\text{n},{{\mathbf{k}}_{y}}}{{{\mathcal{E}}_{\text{N},\text{n},{{\mathbf{k}}_{y}}-{e\mathbf{A}\left( t \right)}/{\hbar }\;}}c_{\text{N},\text{n},{{\mathbf{k}}_{y}}}^{\dagger }{{c}_{\text{N},\text{n},{{\mathbf{k}}_{y}}}}}+\sum\limits_{\mathbf{q}}{\hbar {{\omega }_{\mathbf{q}}}\left( b_{\mathbf{q}}^{\dagger }{{b}_{\mathbf{q}}}+\dfrac{1}{2} \right)},\label{ha2} \\ 
  {{\mathcal{H}}_{\operatorname{int}}}&=\sum\limits_{\text{n},{{\text{n}}^{\prime }}}{\sum\limits_{\text{N},{{\text{N}}^{\prime }}}{\sum\limits_{\mathbf{q},{{\mathbf{k}}_{y}}}{\mathcal{C}\left( \mathbf{q} \right){{\mathcal{I}}_{\text{n},{{\text{n}}^{\prime }}}}\left( {{q}_{z}} \right){{\mathcal{J}}_{\text{N},{{\text{N}}^{\prime }}}}\left( {{\mathbf{q}}_{\bot }} \right)c_{\text{N},\text{n},{{\mathbf{k}}_{y}}+{{\mathbf{q}}_{y}}}^{\dagger }{{c}_{\text{N},\text{n},{{\mathbf{k}}_{y}}}}\left( {{b}_{\mathbf{q}}}+b_{-\mathbf{q}}^{\dagger } \right)}}}, \label{ha3}
\end{align}
here, $c_{\text{N},\text{n},{{\mathbf{k}}_{y}}}^{\dagger }\left( b_{\mathbf{q}}^{\dagger } \right)$, and ${{c}_{\text{N},\text{n},{{\mathbf{k}}_{y}}}}\left( {{b}_{\mathbf{q}}} \right)$ are the creation and the annihilation operators of electron (phonon). $\hbar {{\omega }_{\mathbf{q}}}=\hbar {{\upsilon }_{\text{S}}}q$ is the acoustic phonon energy, with ${{\upsilon }_{\text{S}}}$ is the speed of sound, and $q\equiv \left| \mathbf{q} \right|=\sqrt{\mathbf{q}_{\bot }^{2}+\mathbf{q}_{z}^{2}}$ is the magnitude of the acoustic phonon wave vector. $\mathbf{A}\left( t \right)=\left( 0,{{{E}_{0}}\cos \Omega \text{t}}/{\Omega }\;,0 \right)$ is the vector potential of IEMW in the dipole approximation \cite{mmlc, mmlc2}. $\mathcal{C}\left( \mathbf{q} \right)$ is the electron-acoustic phonon matrix element 
\begin{align}
    {{\left| \mathcal{C}\left( \mathbf{q} \right) \right|}^{2}}=\dfrac{\hbar \mathcal{D}_{\text{ac}}^{2}q}{2\rho {{\upsilon }_{\text{S}}}{{V}_{0}}},
\end{align}
with ${{\mathcal{D}}_{\text{ac}}}$, $\rho$, and ${{V}_{0}}={{L}_{x}}{{L}_{y}}L$ are the acoustic deformation potential, mass density, and normalization volume of specimen. ${{\mathcal{I}}_{\text{n},{{\text{n}}^{\prime }}}}\left( {{q}_{z}} \right)=\left\langle  {{\phi }_{\text{n}}}\left( z \right) \right|{{\mathrm{e}}^{\pm i{{q}_{z}}z}}\left| {{\phi }_{{{\text{n}}^{\prime }}}}\left( z \right) \right\rangle$ is the electron form factor in the confinement potential the infinite semi-parabolic plus semi-inverse squared quantum wells. ${\mathcal{J}}_{\mathrm{N,N^{\prime}}}\left({\mathbf{q}}_{\bot}\right)$ is the magnetic form factor similar to previous studies \cite{hsr1,jnn}, we have 
\begin{align}
    {{\left| {{\mathcal{J}}_{\text{N},{{\text{N}}^{\prime }}}}\left( {{\mathbf{q}}_{\bot }} \right) \right|}^{2}}={{\left| \left\langle  {{\Phi }_{\text{N}}}\left( x-{{x}_{0}} \right) \right|{{e}^{\pm i{{q}_{x}}x}}\left| {{\phi }_{{{\text{N}}^{\prime }}}}\left( x-{{x}_{0}} \right) \right\rangle  \right|}^{2}}=\dfrac{\text{N}!}{{{\text{N}}^{\prime }}!}{{\text{e}}^{-{{u}}}}u^{{{\text{N}}^{\prime }}-\text{N}}\mathcal{L}_{\text{N}}^{{{\text{N}}^{\prime }}-\text{N}}\left( {{u}} \right),
\end{align}
with ${{u}}={\mathbf{q}_{\bot }^{2}\ell _{B}^{2}}/{2}\;$. 

In the relaxation time approximation, the quantum kinetic equation for electrons may be written in the general form as
\begin{align}\label{ptd1}
    \dfrac{\partial {{\mathcal{F}}_{\text{N},\text{n},{{\mathbf{k}}_{y}}}}\left( t \right)}{\partial t}-\left[ e\mathbf{E}+\dfrac{e\hbar }{{{m}_{e}}}\left( {{\mathbf{k}}_{y}}\times \mathbf{B} \right) \right]\dfrac{\partial {{\mathcal{F}}_{\text{N},\text{n},{{\mathbf{k}}_{y}}}}\left( t \right)}{\hbar \partial {{\mathbf{k}}_{y}}}=-\dfrac{{{\mathcal{F}}_{\text{N},\text{n},{{\mathbf{k}}_{y}}}}\left( t \right)-{\mathcal{F}_{\text{N},\text{n},{{\mathbf{k}}_{y}}}^{\left( 0 \right)}}}{\tau },
\end{align}
where, ${{\mathcal{F}}_{\text{N},\text{n},{{\mathbf{k}}_{y}}}}\left( t \right)={{\left\langle c_{\text{N},\text{n},{{\mathbf{k}}_{y}}}^{\dagger }{{c}_{\text{N},\text{n},{{\mathbf{k}}_{y}}}} \right\rangle }_{t}}$ is an unknown distribution function perturbed due to the external fields, and $\tau$ is the constant electron momentum relaxation time. ${{\left\langle X \right\rangle }_{t}}=\text{Tr}\left( \widehat{\rho }\widehat{X} \right)$ denotes a statistical average value at the moment t, $\widehat{\rho }$ is the density matrix operator, the summation over the diagonal elements of the matrix, $\widehat{\rho }\widehat{X}$, or the trace is labeled by Tr. If the temperature is low enough, the electrons system is degenerate and the equilibrium electron distribution function can be assumed to be the Heaviside step function \cite{hpb} ${\mathcal{F}_{\text{N},\text{n},{{\mathbf{k}}_{y}}}^{\left( 0 \right)}}=\theta \left( {{\mathcal{E}}_{\text{F}}}-{{\mathcal{E}}_{\text{N},\text{n},{{\mathbf{k}}_{y}}}} \right)$, in which, ${{\mathcal{E}}_{\text{F}}}$ being the Fermi energy. The first term in the left-hand side of Eq. \eqref{ptd1} characterizes the time evolution of the distribution function perturbed by the teraherz laser field, found using the quantum kinetic equation as in previous studies \cite{hsr1, qke1, qke3}
\begin{align}\label{ptd2}
    \dfrac{\partial {{\mathcal{F}}_{\text{N},\text{n},{{\mathbf{k}}_{y}}}}\left( t \right)}{\partial t}=-\dfrac{i}{\hbar }{{\left\langle \left[ c_{\text{N},\text{n},{{\mathbf{k}}_{y}}}^{\dagger }{{c}_{\text{N},\text{n},{{\mathbf{k}}_{y}}}},{{\mathcal{H}}_{0}}+{{\mathcal{H}}_{\operatorname{int}}} \right] \right\rangle }_{t}}.
\end{align}

Substituting the Hamiltonian from Eqs. \eqref{ha1}, \eqref{ha2}, and \eqref{ha3} into Eq. \eqref{ptd2}, then performing the operator algebraic transformations, we can rewrite Eq. \eqref{ptd1} as
\begin{align}\label{ptdtq}
 \begin{split}
      & -\left[ e\mathbf{E}+\dfrac{e\hbar }{{{m}_{e}}}\left( {{\mathbf{k}}_{y}}\times \mathbf{B} \right) \right]\dfrac{\partial {{\mathcal{F}}_{\text{N},\text{n},{{\mathbf{k}}_{y}}}}\left( t \right)}{\hbar \partial {{\mathbf{k}}_{y}}}=-\dfrac{{{\mathcal{F}}_{\text{N},\text{n},{{\mathbf{k}}_{y}}}}\left( t \right)-{\mathcal{F}_{\text{N},\text{n},{{\mathbf{k}}_{y}}}^{\left( 0 \right)}}}{\tau } \\ 
 & +\dfrac{2\pi }{\hbar }\sum\limits_{{{\text{n}}^{\prime }},\mathbf{q}}{{{\left| \mathcal{C}\left( \mathbf{q} \right) \right|}^{2}}{{\left| {{\mathcal{I}}_{\text{n},{{\text{n}}^{\prime }}}}\left( {{q}_{z}} \right) \right|}^{2}}{{\left| {{\mathcal{J}}_{\text{N},{{\text{N}}^{\prime }}}}\left( {{\mathbf{q}}_{\bot }} \right) \right|}^{2}}}\sum\limits_{\ell =-\infty }^{+\infty }{\text{J}_{\ell }^{2}}\left( \mathcal{D}{{q}_{y}} \right) \\ 
  & \times \left\{ \left[ {{\mathcal{F}}_{{{\text{N}}^{\prime }},{{\text{n}}^{\prime }},{{\mathbf{k}}_{y}}+{{\mathbf{q}}_{y}}}}\left( {{\mathcal{N}}_{\mathbf{q}}}+1 \right)-{{\mathcal{F}}_{\text{N},\text{n},{{\mathbf{k}}_{y}}}}{{\mathcal{N}}_{\mathbf{q}}} \right] \right.\delta \left( {{\mathcal{E}}_{{{\text{N}}^{\prime }},{{\text{n}}^{\prime }},{{\mathbf{k}}_{y}}+{{\mathbf{q}}_{y}}}}-{{\mathcal{E}}_{\text{N},\text{n},{{\mathbf{k}}_{y}}}}-\hbar {{\omega }_{\mathbf{q}}}-\ell \hbar \Omega  \right) \\ 
 & +\left[ {{\mathcal{F}}_{{{\text{N}}^{\prime }},{{\text{n}}^{\prime }},{{\mathbf{k}}_{y}}-{{\mathbf{q}}_{y}}}}{{\mathcal{N}}_{\mathbf{q}}} \right.\left. \left. -{{\mathcal{F}}_{\text{N},\text{n},{{\mathbf{k}}_{y}}}}\left( {{\mathcal{N}}_{\mathbf{q}}}+1 \right) \right]\delta \left( {{\mathcal{E}}_{{{\text{N}}^{\prime }},{{\text{n}}^{\prime }},{{\mathbf{k}}_{y}}-{{\mathbf{q}}_{y}}}}-{{\mathcal{E}}_{\text{N},\text{n},{{\mathbf{k}}_{y}}}}+\hbar {{\omega }_{\mathbf{q}}}-\ell \hbar \Omega  \right) \right\},
 \end{split}
\end{align}
here, ${\mathrm{J}}_{\ell}\left(x\right)$ is the $\ell$th-order Bessel function of the argument $x$. $\mathcal{D}={e{{E}_{0}}}/{\left( {{m}_{e}}{{\Omega }^{2}} \right)}\;$ is the laser dressing parameter.  ${{\mathcal{N}}_{\mathbf{q}}}$ is
the equilibrium distribution function for phonons, which is given by the
Bose–Einstein distribution function. Additionally, we may further simplify the calculation process via using the high-temperature approximation for ${{\mathcal{N}}_{\mathbf{q}}}$ \cite{van1}, ${{\mathcal{N}}_{\mathbf{q}}}\approx {{\mathcal{N}}_{\mathbf{q}}}+1\approx {{{k}_{B}}T}/{\left( \hbar {{\omega }_{\mathbf{q}}} \right)}\;={{{k}_{B}}T}/{\left( \hbar {{\upsilon }_{\text{S}}}q \right)}\;$, with $k_B$ being the Boltzmann constant. Dirac delta functions $\delta\left(x\right)$ with argument $x$  appear due to the energy-momentum conservation law or also called the selection rule. Under the condition that electron-acoustic phonon scattering is elastic, the phonon energy is much smaller than the photon energy and cyclotron energy, so we can ignore the acoustic phonon energy in the argument of the Dirac delta functions \cite{van2,van1}. 
\subsection{Analytical expressions of the conductivity tensors, and longitudinal magnetoresistance.}
To calculate the longitudinal magnetoresistance, we need to know the analytical expression of the longitudinal and Hall conductivities. On the other hand, in the linear response theory, Ohm's law states that the relationship between the total current density ${{\bf{J}}_i} = \sum\limits_{i,j = x,y} {{\sigma _{ij}}{{\bf{E}}_j}}$, in which, ${{\sigma _{xx}}}$, and ${{\sigma _{xy}}}$ are the longitudinal and Hall conductivities; ${{\bf{E}}_x} = {{\bf{E}}_c}$ is the crossed electric field, and ${{\bf{E}}_y} = {{\bf{E}}_{\rm{Hall}}}$ being the induced electric field (or called the Hall field). The longitudinal magnetoresistance is defined via the conductivity tensor as follow \cite{mr1,mr2,qke2,qke1}
\begin{align}
   {\rho _{xx}} &= \dfrac{{{\sigma _{xx}}}}{{\sigma _{xx}^2 + \sigma _{yx}^2}},\label{rxx}
\end{align}

Now, we will briefly present the process of solving Eq. \eqref{ptdtq} to find the expression for the total current density. First, we have the definition of the total current density \cite{qke4}
\begin{align}\label{jt}
  \mathbf{J}\left( t \right)=\int\limits_{0}^{+\infty }{\mathbf{j}\left( \mathcal{E} \right)d\mathcal{E}}=\dfrac{e\hbar }{{{m}_{e}}}\int\limits_{0}^{+\infty }{\sum\limits_{\text{N},\text{n},{{\mathbf{k}}_{y}}}{{{\mathbf{k}}_{y}}{{\mathcal{F}}_{\text{N},\text{n},{{\mathbf{k}}_{y}}}}\left( t \right)\delta \left( \mathcal{E}-{{\mathcal{E}}_{\text{N},\text{n},{{\mathbf{k}}_{y}}}} \right)d\mathcal{E}}},
\end{align}
with $\mathbf{j}\left( \mathcal{E} \right)=\dfrac{e\hbar }{{{m}_{e}}}\sum\limits_{\text{N},\text{n},{{\mathbf{k}}_{y}}}{{{\mathbf{k}}_{y}}{{\mathcal{F}}_{\text{N},\text{n},{{\mathbf{k}}_{y}}}}\left( t \right)}$ is the partial current density (the current caused by electrons that have energy of $\mathcal{E}$). Next, to find $\mathbf{j}\left( \mathcal{E} \right)$, we multiply both sides of Eq. \eqref{ptdtq} by $\dfrac{e\hbar }{{{m}_{e}}}{{\mathbf{k}}_{y}}\delta \left( \mathcal{E}-{{\mathcal{E}}_{\text{N},\text{n},{{\mathbf{k}}_{y}}}} \right)$ and then sum over $\text{N},\text{n},{{\mathbf{k}}_{y}}$. For simplicity, we limit the problem to considering only one-photon absorption and emission processes. That also means that the order of the Bessel function in Eq. \eqref{ptdtq} only considers the values $\ell = -1, 0,\text{ and 1}$. After straightforward calculations, we find the quantum kinetic equation for $\mathbf{j}\left( \mathcal{E} \right)$ as follows \cite{mr1,mr2}
\begin{align}\label{ptjt}
    \dfrac{\mathbf{j}\left( \mathcal{E} \right)}{\tau }+\dfrac{e}{{{m}_{e}}}\left[ \mathbf{B}\times \mathbf{j}\left( \mathcal{E} \right) \right]=\mathbf{Q}\left( \mathcal{E} \right)+\mathbf{S}\left( \mathcal{E} \right),
\end{align}
where, 
\begin{align}
    \mathbf{Q}\left( \mathcal{E} \right)=-\dfrac{e\hbar }{{{m}_{e}}}\sum\limits_{\text{N},\text{n},{{\mathbf{k}}_{y}}}{{{\mathbf{k}}_{y}}\cdot \left( e{{\mathbf{E}}_{c}}\cdot \dfrac{\partial {{\mathcal{F}}_{\text{N},\text{n},{{\mathbf{k}}_{y}}}}}{\hbar \partial {{\mathbf{k}}_{y}}} \right)\delta \left( \mathcal{E}-{{\mathcal{E}}_{\text{N},\text{n},{{\mathbf{k}}_{y}}}} \right),}
\end{align}
\begin{align}
\begin{split}
       \mathbf{S}\left( \mathcal{E} \right)&=\dfrac{4\pi e}{{{m}_{e}}}\sum\limits_{{{\text{N}}^{\prime }},{{\text{n}}^{\prime }},\text{N},\text{n}}{\sum\limits_{\mathbf{q},{{\mathbf{k}}_{y}}}{{{\left| \mathcal{C}\left( \mathbf{q} \right) \right|}^{2}}{{\left| {{\mathcal{I}}_{\text{n},{{\text{n}}^{\prime }}}}\left( {{q}_{z}} \right) \right|}^{2}}{{\left| {{\mathcal{J}}_{\text{N},{{\text{N}}^{\prime }}}}\left( {{\mathbf{q}}_{\bot }} \right) \right|}^{2}}{{\mathcal{N}}_{\mathbf{q}}}{{\mathbf{k}}_{y}}\delta \left( \mathcal{E}-{{\mathcal{E}}_{\text{N},\text{n},{{\mathbf{k}}_{y}}}} \right)\left( {{\mathcal{F}}_{{{\text{N}}^{\prime }},{{\text{n}}^{\prime }},{{\mathbf{k}}_{y}}+{{\mathbf{q}}_{y}}}}-{{\mathcal{F}}_{\text{N},\text{n},{{\mathbf{k}}_{y}}}} \right)}} \\ 
 & \times \left[ \left( 1-\dfrac{{{\mathcal{D}}^{2}}q_{y}^{2}}{2} \right)\delta \left( {{\mathcal{E}}_{{{\text{N}}^{\prime }},{{\text{n}}^{\prime }},{{\mathbf{k}}_{y}}+{{\mathbf{q}}_{y}}}}-{{\mathcal{E}}_{\text{N},\text{n},{{\mathbf{k}}_{y}}}} \right)+\dfrac{{{\mathcal{D}}^{2}}q_{y}^{2}}{4}\delta \left( {{\mathcal{E}}_{{{\text{N}}^{\prime }},{{\text{n}}^{\prime }},{{\mathbf{k}}_{y}}+{{\mathbf{q}}_{y}}}}-{{\mathcal{E}}_{\text{N},\text{n},{{\mathbf{k}}_{y}}}}\pm \hbar \Omega  \right) \right].  
\end{split}
\end{align}

Solving Eq. \eqref{ptjt} similarly to previous studies \cite{mr1}, we obtain 
\begin{align}\label{pjt}
    \mathbf{j}\left( \mathcal{E} \right)=\dfrac{\tau }{1+\omega _{B}^{2}{{\tau }^{2}}}\left\{ \mathbf{Q}\left( \mathcal{E} \right)+\mathbf{S}\left( \mathcal{E} \right)-\dfrac{e\tau }{{{m}_{e}}}\mathbf{B}\times \left[ \mathbf{Q}\left( \mathcal{E} \right)+\mathbf{S}\left( \mathcal{E} \right) \right] \right\}.
\end{align}

Inserting Eq. \eqref{pjt} into Eq. \eqref{jt} and doing some mathematical manipulation with
the transformations and Poisson's summation formula  \cite{van1, van2}, we obtain the expression for the the longitudinal and Hall conductivities 
\begin{align}
   {{\sigma }_{xx}}&=\dfrac{\tau }{1+\omega _{B}^{2}{{\tau }^{2}}}\left( a+\dfrac{eb}{{{m}_{e}}}\dfrac{1-\omega _{B}^{2}{{\tau }^{2}}}{1+\omega _{B}^{2}{{\tau }^{2}}} \right), \label{sigxx}\\ 
  {{\sigma }_{yx}}&=\dfrac{{{\omega }_{B}}{{\tau }^{2}}}{1+\omega _{B}^{2}{{\tau }^{2}}}\left( a+\dfrac{eb}{{{m}_{e}}}\dfrac{2\tau }{1+\omega _{B}^{2}{{\tau }^{2}}} \right),\label{sigyx} \\ 
  a&=\dfrac{{{\eta }_{0}}{{e}^{2}}{{L}_{y}}}{2\pi {{m}_{e}}\hbar {{\upsilon }_{d}}}\sum\limits_{\text{N},\text{n}}{\left( {{\mathcal{E}}_{\text{N},\text{n}}}-{{\mathcal{E}}_{\text{F}}} \right),}
\end{align}
\begin{align}\label{bco}
    \begin{split}
 b&={{b}_{0}}\left\{ \left( 1-\dfrac{{{\mathcal{D}}^{2}}{{\overline{\ell }}^{2}}}{2\ell _{B}^{4}} \right)\left[ 1+2\sum\limits_{\kappa =1}^{\infty }{{{\left( -1 \right)}^{\kappa }}{{\text{e}}^{-{2\pi \kappa \Gamma }/{\left( \hbar {{\omega }_{B}} \right)}\;}}\text{cos}\left( 2\pi \kappa \aleph  \right)} \right] \right. \\ 
 & \left. +\dfrac{{{\mathcal{D}}^{2}}{{\overline{\ell }}^{2}}}{4\ell _{B}^{4}}\left[ 1+2\sum\limits_{\kappa =1}^{\infty }{{{\left( -1 \right)}^{\kappa }}{{\text{e}}^{-{2\pi \kappa \Gamma }/{\left( \hbar {{\omega }_{B}} \right)}\;}}\text{cos}\left( 2\pi \kappa {{\aleph }_{\pm }} \right)} \right] \right\},
    \end{split}
\end{align}
\begin{align}
      {{b}_{0}}& =\dfrac{e{{\eta }_{0}}\mathcal{D}_{\text{ac}}^{2}{{m}_{e}}{{k}_{B}}T{{L}_{y}}L}{8{{\pi }^{2}}{{\hbar }^{4}}\rho {{\upsilon }_{d}}\upsilon _{\text{S}}^{3}}{{\text{G}}_{\text{n},{{\text{n}}^{\prime }}}}\bar{\ell }\left( {{\mathcal{E}}_{\text{N},\text{n}}}-{{\mathcal{E}}_{\text{F}}} \right),\\ 
  {{\text{G}}_{\text{n},{{\text{n}}^{\prime }}}}&=\int\limits_{-\infty }^{+\infty }{{{\left| {{\mathcal{I}}_{\text{n},{{\text{n}}^{\prime }}}}\left( {{q}_{z}} \right) \right|}^{2}}d{{q}_{z}}}, 
\end{align}
here, ${{\text{G}}_{\text{n},{{\text{n}}^{\prime }}}}$ is the squared overlap integral describing crystal momentum conservation \cite{ampli2}, $\eta_0$ is the electron density, and $\Gamma ={\hbar }/{\tau }\;$ is the damping factor. The appearance of the parameter $\overline{\ell }={\left( \sqrt{\text{N}+{1}/{2}\;}+\sqrt{\text{N}+{3}/{2}\;} \right){{\ell }_{B}}}/{2}\;$ is derived from the assumption that there exists an effective phonon momentum such that $e{{\upsilon }_{d}}{{q}_{y}}\approx eE\overline{\ell }$ by C. M. Van Vliet \textit{et al.} \cite{van3}. ${\aleph }$, and ${{\aleph }_{\pm }}$ appear when transforming the Dirac delta function from the Poisson’s summation formula as in \cite{van2, van1, van3} 
\begin{align}\label{nct}
   \aleph &=\dfrac{{{\mathcal{E}}_{{{\text{N}}^{\prime }},{{\text{n}}^{\prime }}}}-{{\mathcal{E}}_{\text{N},\text{n}}}+eE\bar{\ell }}{\hbar {{\omega }_{B}}}, \\ 
 & {{\aleph }_{\pm }}=\aleph \pm \dfrac{\Omega }{{{\omega }_{B}}}, \label{nct1}
\end{align}
with ${{\mathcal{E}}_{\text{N},\text{n}}}$ in Eq. \eqref{enn}. The $+/-$ signs characterize the one-photon emission/absorption process. We note that the ratio ${\Omega }/{{{\omega }_{B}}}\;$ appears in Eq. \eqref{nct1} governs the period of the regular Shubnikov–de Haas oscillations has been well known in previous study \cite{qhemw,qhemw2}. 

Finally, putting the results from Eqs. \eqref{sigxx} and \eqref{sigyx} into \eqref{rxx}, we obtain a general analytical expression for the longitudinal magnetoresistance in the infinite semi-parabolic plus semi-inverse squared quantum wells under the influence of IEMW. We can see the complex dependence of longitudinal magnetoresistance on the geometric structural parameters of the quantum well and the external fields. The results include calculations in both the asymmetric semi-parabolic ($\beta_z = 0$) and the asymmetric semi-parabolic plus inverse squared ($\beta_z \ne 0$) quantum well cases. In other confined models such as parabolic \cite{mr1} and rectangular \cite{mr3} potential quantum
wells, the difference in the confinement structure leads to changes in the wave function and energy spectrum, which will lead to new results on the electronic form factor, longitudinal magnetoresistance. In the next section, to provide further insights, we will present detailed numerical results with the help of computer programs. 

\section{Results and Discussions}
To clarify physical meanings of the above obtained result, in this section, we numerically evaluate the the longitudinal magnetoresistance and Hall coefficient using specific parameters of $\text{GaAs}/\text{AlGaAs}$ heterostructures. The parameters taken in the evaluation are given in the Tab. \ref{tab1}. In this paper, we only consider transitions that occur between the lowest neighboring energy levels and ignore the considered small contributions from transitions between distant states. Therefore, we only consider the electron transitions between the lowest subbands $\text{n}=0\to {{\text{n}}^{\prime }}=1,\text{N}=0 \to {{\text{N}}^{\prime }}=1$. 

\begin{table}[!htb]
\centering
\caption{Main input parameters of our calculation model.}
\begin{tabular}{clcc}
\toprule
\hline
\multicolumn{1}{c}{Symbol} & \multicolumn{1}{c}{Definition} & Value & Unit \\ \hline
$\eta_0$ & Electron density \cite{mr1,fqw1}  & $3.0 \times 10^{16}$ & $\text{cm}^{-3}$ \\
$\rho$ & Mass density \cite{exp0,exp1,exp3}  & $5.37$ & $\text{g/cm}^3$ \\
${{\upsilon }_{\text{S}}}$ & Speed of sound \cite{exp0, exp1,exp3}  & $5.22\times {{10}^{3}}$ & $\text{m/s}$ \\
${{\mathcal{D}}_{\text{ac}}}$ & Deformation potential constant \cite{exp0,exp1,exp3}  & 12.42 & $\text{eV}$ \\
$\tau$ & Constant relaxation time \cite{exp2} & $1.0 \times {10^{ - 12}}$ & s \\
\begin{tabular}[c]{@{}l@{}}${{{m_e}} \mathord{\left/
 {\vphantom {{{m_e}} {{m_0}}}} \right.
 \kern-\nulldelimiterspace} {{m_0}}}$\end{tabular} & \begin{tabular}[c]{@{}l@{}}Ratio of the effective mass \\ to the rest mass of electrons \cite{exp0,exp1,hsr1,exp3}\end{tabular} & 0.067 &  \\
${\mathcal{E}_{\text{F}}}$ & Fermi energy \cite{hsr1}& $50$ & ${\text{meV}}$ \\ 
${E_{x}}$ & Crossed electric field \cite{mr1}& $5$ & ${\text{V/cm}}$ \\ 
\hline
\bottomrule
\end{tabular}
\label{tab1}
\end{table}

\begin{figure}[!htb]
\centering
\subfigure[][ \label{2a}]
  {\includegraphics[scale=0.5]{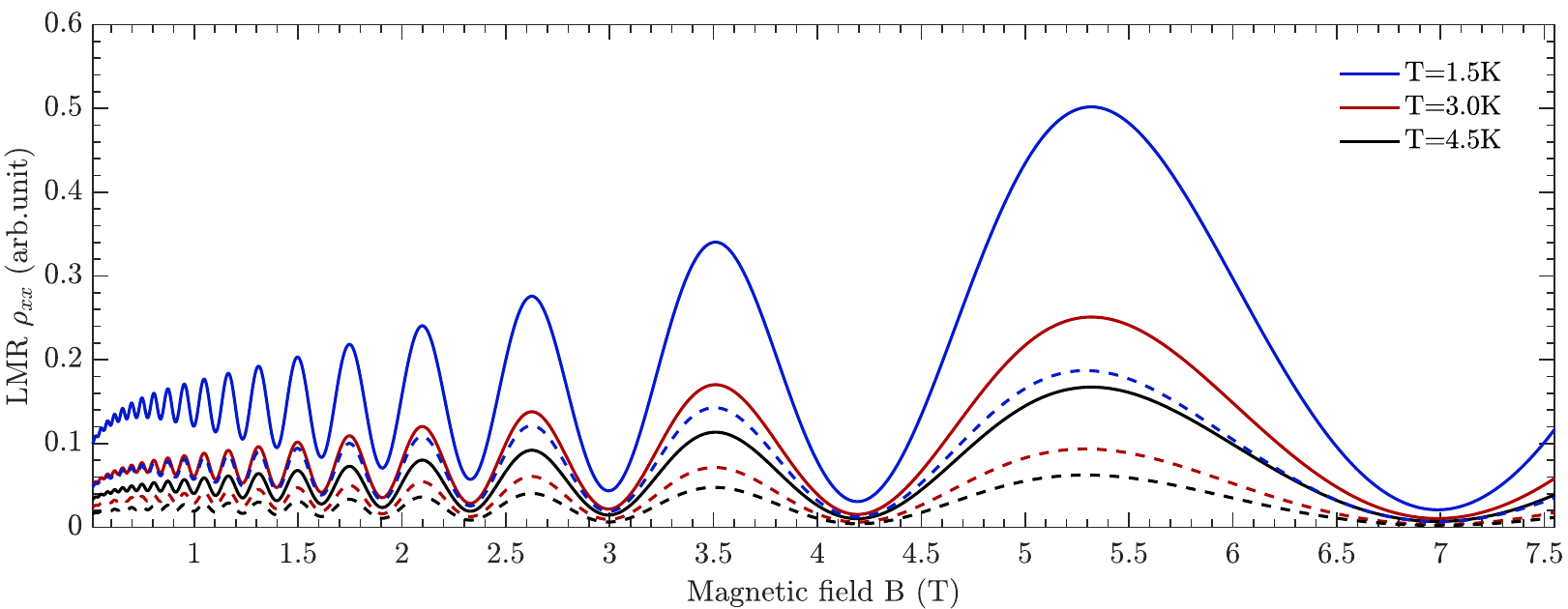}}
\subfigure[][ \label{2b}]
  {\includegraphics[scale=0.5]{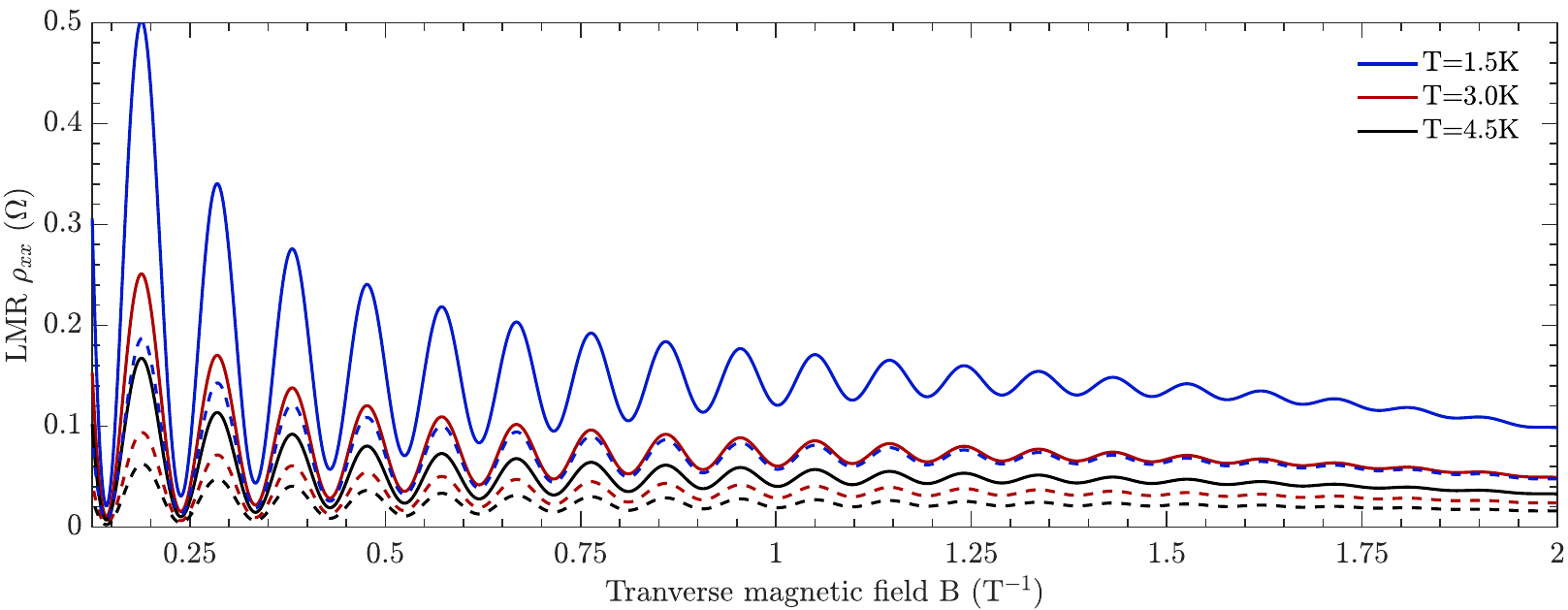}}
\caption{The variation of the LMR 
 with (a) the magnetic field and (b) inverse magnetic field for the semi-parabolic (dashed-line) and semi-parabolic plus semi-inverse squared (solid-line) quantum wells  under the three various temperatures where the blue-line, red-line, black-line, correspond to the T = 1.5 K, 3.0 K, and 4.5 K. Here, $\omega_z = 55$THz \cite{mr1}.} 
\label{fig2}
\end{figure}

In Fig. \ref{fig2}, we show the LMR as a function of the magnetic field (Fig. \ref{2a}) and inverse magnetic field (Fig. \ref{2b}) at different values of the temperature in the absence of IEMW. It can be seen that temperature only affects the oscillation amplitude but does not change the position of the maximum and minimum peaks. Specifically, as the temperature increases, the amplitude of oscillations tends to decrease. This will be discussed in detail later. Now, we would like to draw the reader's attention to the change in the period of oscillations due to the increase in magnetic field strength. In Fig. \ref{2b}, it can be seen that the period of the SdH oscillations is proportional to the inverse of the magnetic field. This can be explained quantitatively from Eq. \eqref{bco}, where the period of the oscillation is inversely proportional to the cyclotron frequency, on the other hand, the cyclotron frequency is proportional to the magnetic field (recall ${{\omega }_{B}}={eB}/{{{m}_{e}}}\;$), so the dependence of the SdH oscillation period on the magnetic field is according to inverse law (${1}/{B}\;$). This result is consistent with what is known in the parabolic \cite{mr1} and rectangular \cite{mr3} quantum well models. 
Therefore, it can also be concluded that the oscillation period is independent of the geometric structure parameter $\beta_z$ of the quantum wells, which is confirmed in Fig. \ref{2b}. However, from Fig. \ref{2a}, the influence of the geometric structure parameter $\beta_z$ on the LMR and the amplitude of SdH oscillations is significant, especially in the strong magnetic field region. When $\beta_z$ is large, the quantum confinement effect of the quantum well is stronger, especially in the strong magnetic field region, the parabolic potential generated by the magnetic field is also stronger, leading to the sharpness and broadening of the amplitude of the SdH oscillations, which is considered as the characteristic quantum behavior of electrons in the magnetic field at low temperatures. 

\begin{figure}[!htb]
    \centering
    \includegraphics[scale = 0.7]{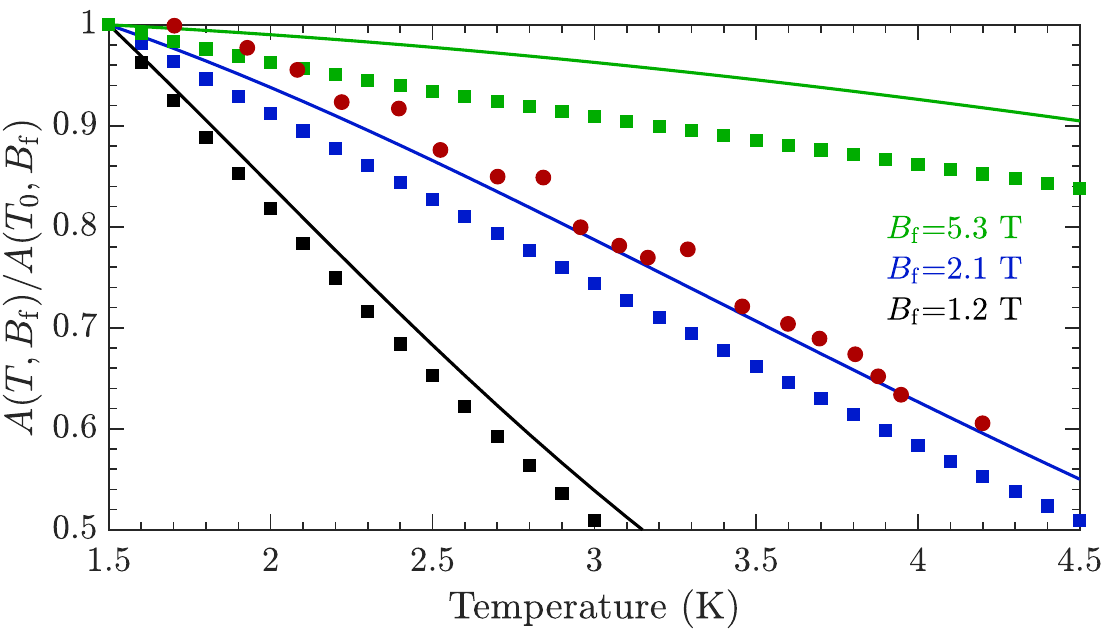}
    \caption{Temperature dependence of the relative amplitude ${A\left( T,{{B}_{\text{f}}} \right)}/{A\left( {{T}_{0}},{{B}_{\text{f}}} \right)}\;$
    defined at different values of the fixed magnetic field $B_{\text{f}}$. The filled squares are our calculation, the red circles are experimental data at $B_{\text{f}} = 2.1$T for multiple quantum wells with aluminum concentration of 0.32 in Ref. \cite{ampli2}, and the solid lines are the theoretical results in Ref. \cite{linke}.} 
    \label{fig3}
\end{figure}

As shown in Fig. \ref{fig2}, the amplitude of the SdH oscillations at a fixed
magnetic field decreases with increasing temperature. We study the monotonically decreasing law of the relative oscillation amplitude at a fixed value of magnetic field quantitatively in Fig. \ref{fig3}. We present the results of relative amplitude calculations at initial temperature $T_0$ = 1.5 K and fixed values of $B_{\text{f}}$ as $B_{\text{f}}$ = 1.2T, 2.1 T and 5.2 T, respectively. We compare the calculation results from our theory with other experimental data of N. Balkan \cite{ampli2} and theoretical prediction (Eq. [12]) in Ref. \cite{linke} as shown in Fig. \ref{fig3}. From the figure, we calculate the relative amplitude values of SdH  oscillations and find that the results obtained by using quantum kinetic equations are in good agreement with previous theoretical and experimental works \cite{linke, ampli2}.

\begin{figure}[!htb]
\centering
\subfigure[][ \label{4a}]
  {\includegraphics[scale=0.5]{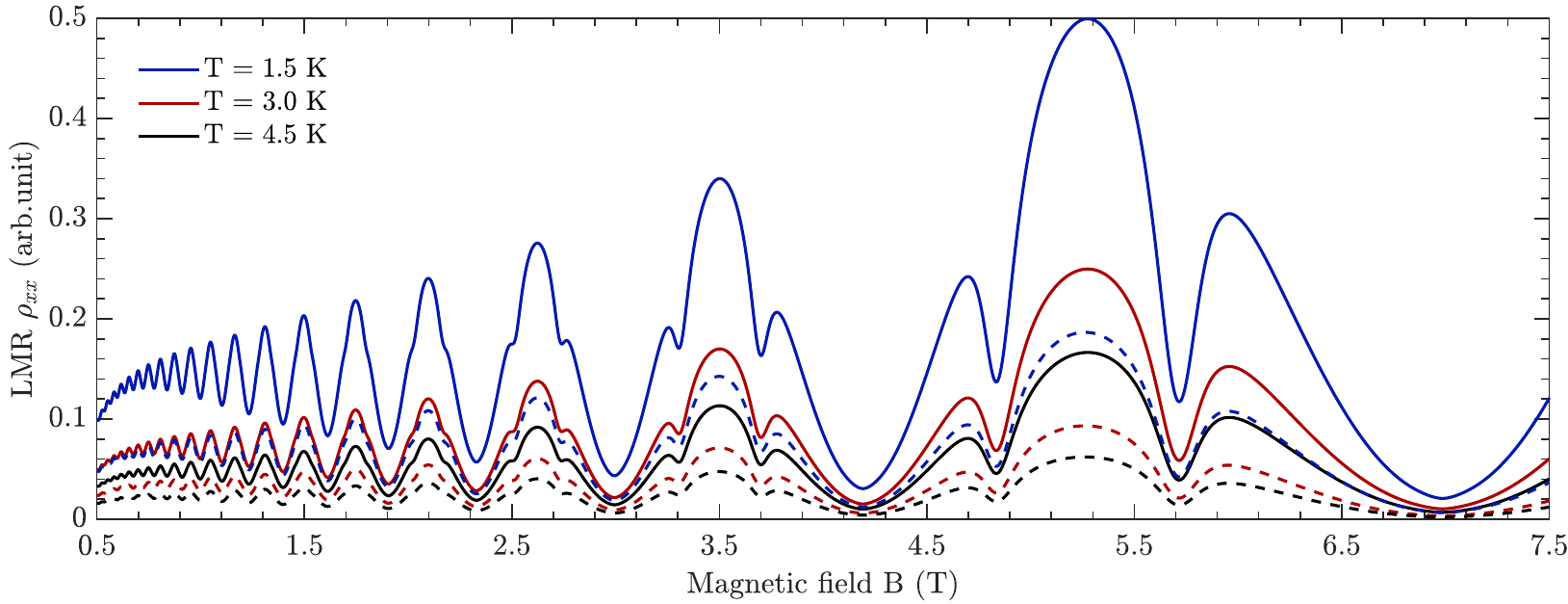}}
\subfigure[][ \label{4b}]
  {\includegraphics[scale=0.5]{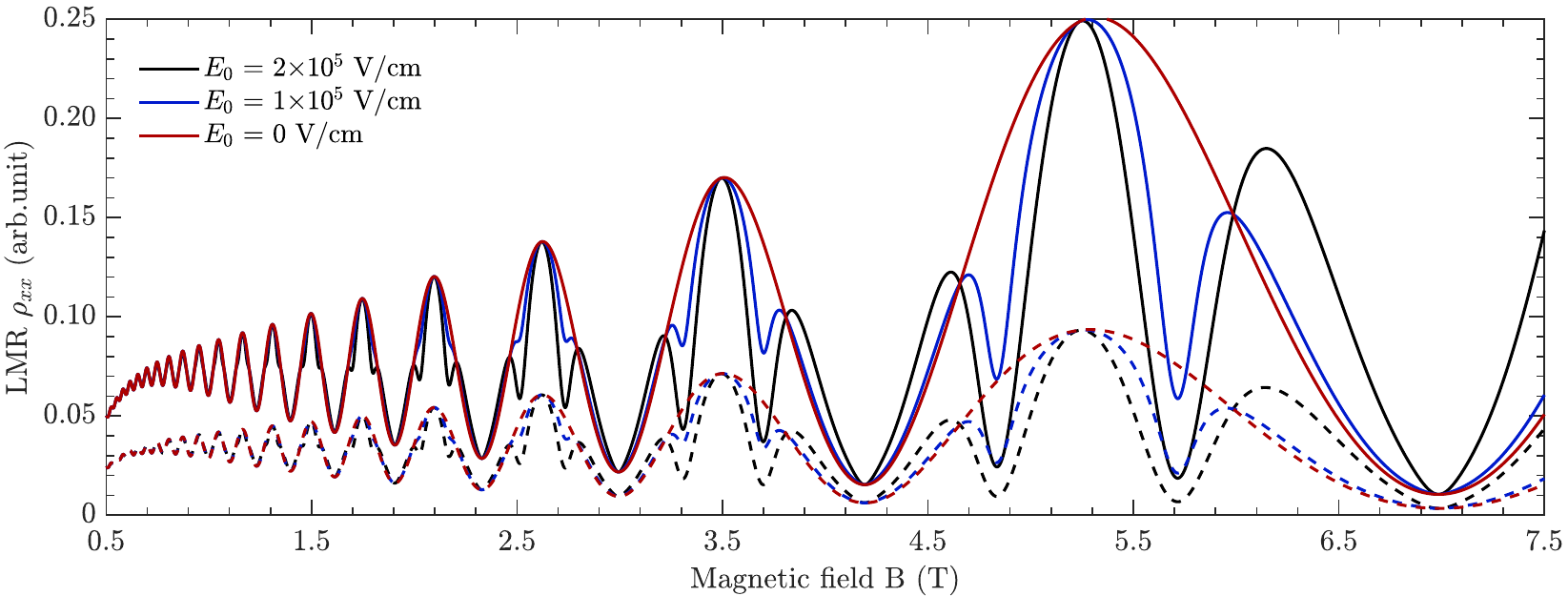}}
\caption{LMR as a function of the magnetic field with asymmetric cases: $\beta_z=0$ (dashed-line) and $\beta_z \ne 0$ (solid-line)  of the semi-parabolic quantum wells at different values of the (a) temperature and (b) intensity of IEMW. Here, $\omega_z = 55$THz. SdH oscillations under the influence of electromagnetic waves are called microwave-induced magnetoresistance oscillations \cite{ampli2}, which is characterized by quantum beat behavior. The quantum beat phenomenon does not appear when there is no strong electromagnetic wave propagating in the system (see Fig. \ref{fig2}).}
\label{fig4}
\end{figure}

The influence of IEMW on the SdH oscillations is shown in Fig. \ref{fig4} where we plot LMR versus B at different values of the temperature with both semi-parabolic  ($\beta_z=0$) and semi-parabolic plus inverse squared ($\beta_z \ne 0$) potential well models. SdH oscillations under the influence of electromagnetic waves are called microwave-induced magnetoresistance oscillations \cite{ampli2}. From Fig. \ref{fig4}, we observe the appearance of quantum-beat phenomenon of SdH oscillations when IEMW propagates in the system. The origin of the beat patterns in SdH oscillations can be simply explained as follows. The energy separation $\Delta \mathcal{E}={{\mathcal{E}}_{{{\text{N}}^{\prime }},{{\text{n}}^{\prime }}}}-{{\mathcal{E}}_{\text{N},\text{n}}}$ that gives rise to magnetoresistance oscillations in Fig. \ref{fig1} is described by the function $\text{cos}\left( 2\pi \kappa \aleph  \right)$ in the first term on the left-hand side of Eq. \eqref{bco}. However, in the presence of IEMW, the characteristic of which is the contribution of the  function $\text{cos}\left( 2\pi \kappa {{\aleph }_{\pm }} \right)$ in the second term on the left-hand side of Eq. \eqref{bco} to the magnetoresistance oscillations.
This is similar to the superposition of oscillations with slightly different frequencies and causes the beat phenomenon as shown in Fig. \ref{fig4}. In addition, also from Fig. \ref{4a}, we also see that the beat phenomenon becomes more obvious in the strong magnetic field region, specifically, with $B > 2.5$T and this value does not depend on the temperature (see Fig. \ref{4a}) as well as the intensity of the IEMW (see Fig. \ref{4b}). This is also explained through the difference in the frequency of the oscillations in the left side of Eq. \eqref{bco}. Indeed, when $B$ increases, $\omega_B$ also increases (recall, ${{\omega }_{B}}={eB}/{{{m}_{e}}}\;$), leading to a decrease in the ratio ${\Omega }/{{{\omega }_{B}}}\;$. As a result, the difference between $\aleph $ and ${{\aleph }_{\pm }}$ is small enough to satisfy the condition for the appearance of beats. Furthermore, it is noteworthy that the structural parameters of the quantum well also do not affect the beats in the SdH oscillation. In other words, the beat phenomenon is an important feature that shows the influence of electromagnetic wave frequency on SdH oscillations in general two-dimensional electron gas models. This is our new contribution, and it can be considered an important criterion in future experimental studies related to SdH oscillations in two-dimensional electron gas systems. 

The dependence on the confinement structure of the potential well (frequency $\omega_z$) of the SdH oscillations is given in Fig. \ref{fig5}. For convenience of observation, we consider this problem in the absence of IEMW (no beats). It can be seen that the SdH oscillations become more obvious as the confinement frequency of the quantum well increases. This is understandable because the appearance of SdH oscillations is a special property of the two-dimensional electronic system under the influence of a magnetic field. When the confinement frequency is small, the two-dimensional electronic system will behave like a three-dimensional electronic system and the SdH oscillations will disappear. This is the same result as that obtained previously in the parabolic quantum well \cite{mr1}.  

In order to investigate in more detail, especially the maxima and minima in the SdH oscillations, we calculate the LMR of the two-dimensional electronic gas in semi-parabolic quantum wells as a function of the ratio of the IEMW frequency to the fixed cyclotron frequency $\omega_B$ with $B$ = 3T. It can be seen that the IEMW intensity does not affect the position of the maximum and minimum peaks of the megnetoresistance oscillations. However, the IEMW intensity strongly affects the amplitude of the oscillations, the greater the IEMW intensity, the stronger the oscillation amplitude. Additionally, the maximum peaks appearing at ${\Omega }/{{{\omega }_{B}}}\;$ are integers, while the minimum peaks appearing at ${\Omega }/{{{\omega }_{B}}}\;$ are half-integers. Similar to the results in Fig. \ref{fig4}, the oscillations are formed by the $\text{cos}\left( 2\pi \kappa \aleph  \right)$ and $\text{cos}\left[ 2\pi \kappa \left( \aleph \pm {\Omega }/{{{\omega }_{B}}}\; \right) \right]$ functions in the left side of Eq. \eqref{bco}. Indeed, from Fig. \eqref{fig6}, with $B$ = 3T, cyclotron frequency $\omega_B = 5.88$THz, and $\omega_z = 55$THz, we get $\aleph \approx 7$. Then, if ${\Omega }/{{{\omega }_{B}}}\;$ are integers, the functions $\text{cos}\left[ 2\pi \kappa \left( \aleph \pm {\Omega }/{{{\omega }_{B}}}\; \right) \right]$ have a maximum value of 1, while if ${\Omega }/{{{\omega }_{B}}}\;$ are half-integers, this function will vanish. Another quantitative explanation was given by O. E. Raichev \cite{raichev} when studying the  dynamic conductivity tensor of a quantum well using the Kubo formalism, assuming elastic scattering of electrons by the random disorder potential with arbitrary correlation length. In addition, the above results are also consistent with previous theoretical study \cite{torres} and experimental observations \cite{qhemw2}. Another point worth noting here is the cyclotron-photon resonance condition, which occurs when the photon frequency is an integer multiple of the cyclotron frequency. Cyclotron resonance plays an important role in studies of energy band structure, electronic states, and interactions between electrons and other quasi-particles in condensed matter physics \cite{in5, in2}. 

Following the reviewer's interesting suggestion, we dedicate the final part of this section to discussing the parabolic limit of the confinement potential $\mathcal{U}(z)$ in this paper by expanding it around the equilibrium point $z = z_{\min}$. Indeed, after some analytic transformations, we obtain
\begin{align}\label{rpw}
\mathcal{U}(z) \approx \mathcal{U}_{\min} (\beta_z, \omega_z) + \frac{1}{2} m_e \Omega_z^2 \left[ z - z_{\min} (\beta_z, \omega_z) \right]^2,
\end{align}
here, $z_{\min} (\beta_z, \omega_z) = \left( \dfrac{\hbar \sqrt{\beta_z}}{m_e \omega_z} \right)^{1/2}, \mathcal{U}_{\min} (\beta_z, \omega_z) = \dfrac{\hbar^2 \beta_z}{2 m_e} + \dfrac{m_e \omega_z^2}{2}$, and $\Omega_z = 2 \omega_z$.

Thus, in the small region around the equilibrium point $z = z_{\min}$, the semi-parabolic plus semi-inverse squared potential can be approximated by a parabolic potential with an effective confinement frequency $\Omega_z = 2 \omega_z$. It is noteworthy that we emphasize the contribution of the asymmetric structural parameter $\beta_z$, which is represented in $\mathcal{U}_{\min}(\beta_z, \omega_z)$ and $z_{\min}(\beta_z, \omega_z)$ of the reduced parabolic potential  in Eq. \eqref{rpw}. In the parabolic quantum well, we have solved the problems related to the quantum Hall effect and the magnetoresistance oscillations by the quantum kinetic equation in both the optical \cite{mr5} and acoustic \cite{mr1} phonon scattering mechanisms. 

\section{Conclusions}
\begin{figure}[!htb]
    \centering
    \includegraphics[scale = 0.5]{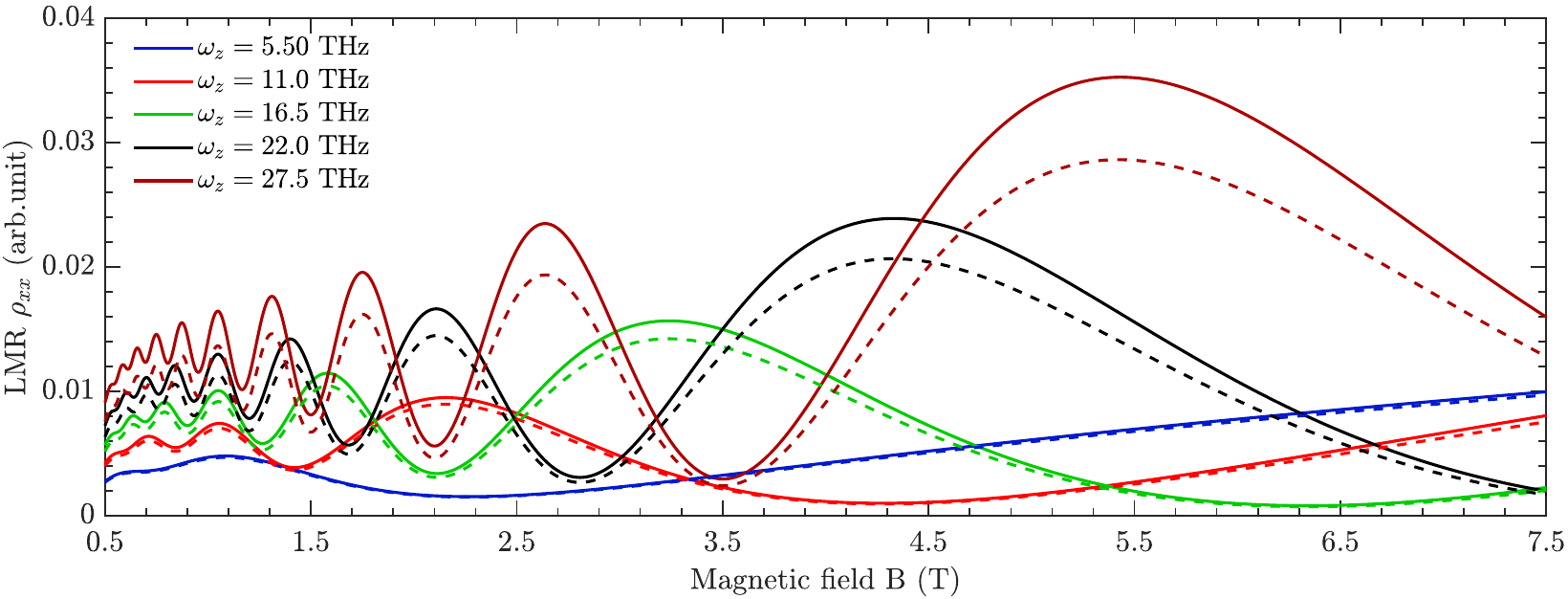}
    \caption{LMR versus the magnetic field for different confinement frequencies at $T=3$K, and $E_0=0$ (without IEMW). The results are presented for the semi-parabolic  (dashed-line) and semi-parabolic plus semi-inverse squared (solid-line) quantum wells. It can be seen that the SdH oscillations become more obvious as the confinement frequency of the quantum well increases.} 
    \label{fig5}
\end{figure}

\begin{figure}[!htb]
    \centering
    \includegraphics[scale = 0.5]{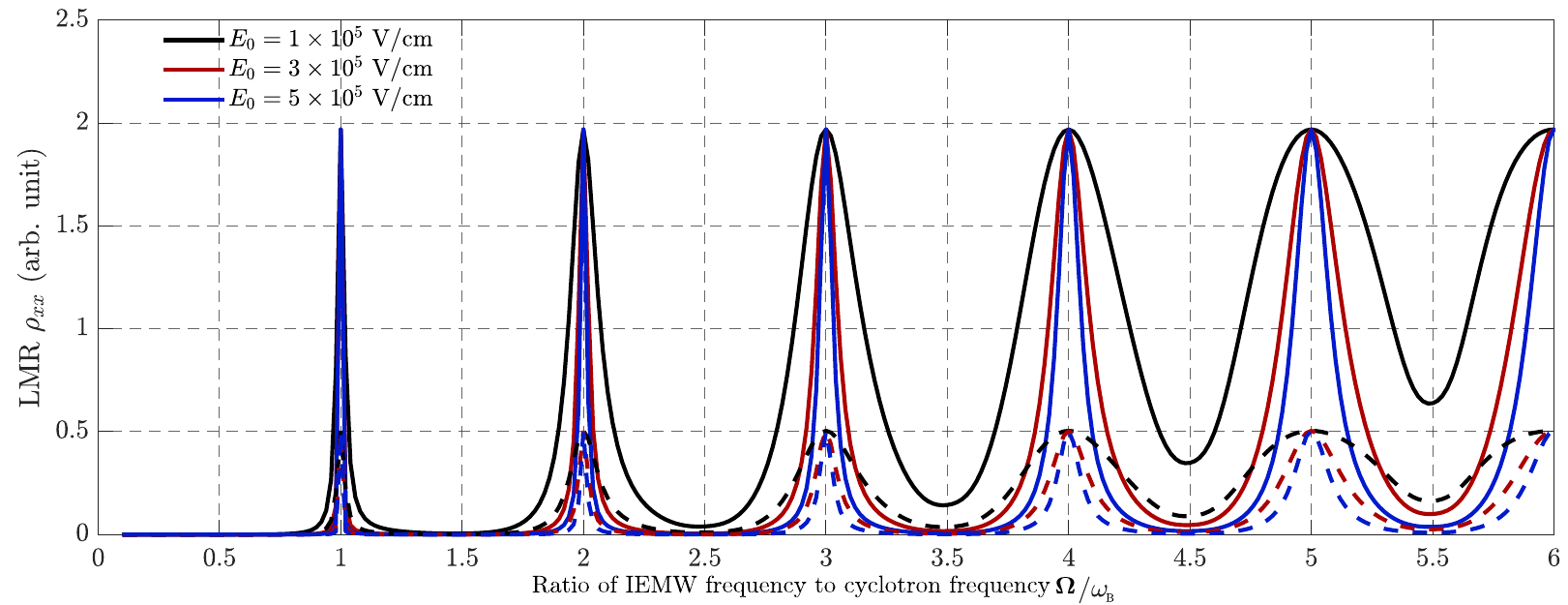}
    \caption{The LMR as functions of the ratio of the IEMW frequency and cyclotron frequency at different values of the intensity of IEMW for the semi-parabolic  (dashed-line) and semi-parabolic plus semi-inverse squared (solid-line) quantum wells. Here, $T=3$K, $\omega_z =55$THz, $B=3$T. The maximum peaks appear to obey the cyclotron resonance condition, i.e. $\Omega =\text{j}{{\omega }_{B}}$, where j is an integer. In addition, the positions of the minimum peaks satisfy $\Omega =\left( \text{j}+{1}/{2}\; \right){{\omega }_{B}}$. These results are in agreement with previous studies by O. E. Raichev \cite{raichev}, M. Torres \cite{torres} and M. Zudov \cite{qhemw2}.}
    \label{fig6}
\end{figure}

The magnetoresistance oscillations of quantum wells described by a infinite asymmetric semi-parabolic confined potential are studied under an intense electromagnetic waves via the electron-acoustic phonon scattering mechanism at low temperatures. We have obtained the analytical expression of LMR for asymmetric cases ($\beta_z=0$ and $\beta_z = 1$) of the semi-parabolic quantum wells using the quantum kinetic equation. Numerical calculations have shown that the LMR depends dramatically on the parameters of external fields such as magnetic field and temperature as well as the structural parameters of the quantum well. In the presence of IEMW, microwave-induced magnetoresistance oscillations appear with beats in the magnetic field region greater than 2.5T with amplitude decreasing with temperature. At the same time, the structural parameters of the quantum well as well as the intensity of IEMW cause an increase in the amplitude of the oscillations. In addition, the influence of electromagnetic wave frequency on LMR is significant due to the formation of maximum and minimum peaks of microwave-induced magnetoresistance oscillations by the ratio of the IEMW frequency to the cyclotron frequency. These results obtained from Figs. \ref{fig2}, \ref{fig4}, \ref{fig5}, and \ref{fig6} show that the amplitude of the magnetoresistance oscillations broadens significantly in the strong magnetic field region and the high frequency region of the intense electromagnetic wave. The appearance of these peaks in Fig. \ref{fig6} corresponds to the cyclotron resonance conditions commonly used in studies of the energy band structure and interactions of electrons with other quasi-particles in crystals. Our results in the limiting cases (absence of IEMW) and zero geometric structure parameters (semi-parabolic quantum wells) are in good agreement with previous theoretical studies and experimental observations. 

In this study, we present the characteristics of magnetoresistance oscillations within the framework of the integer quantum Hall effect in quantum wells with a semi-parabolic plus inverse-square confinement potential. The origin of SdH oscillations arises from the sequential crossing of discrete Landau levels through the Fermi energy at low temperatures. Additionally, in the low magnetic field regime, the role of spin-orbit interaction makes a significant contribution to SdH oscillations in quantum well systems \cite{so1, so2, so3} specifically, as well as in two-dimensional electron systems \cite{so4} in general. In the presence of spin-orbit interaction, the Hamiltonian of the two-dimensional electron system needs to be supplemented with terms characterizing the Rashba and Dresselhaus interactions \cite{so4, so6}. This leads to a dramatic change in the electronic energy spectrum due to  spin splitting and the emergence of beating behavior in SdH oscillations \cite{so6, so5,so7}. Although the contribution of spin-orbit interactions is not taken into account in this paper, the interesting results in the above studies have inspired us to continue applying the quantum kinetic equations approach when considering two-dimensional electronic systems in the presence of spin-orbit interactions in the near future.

\ack
We thank anonymous referees for their valuable comments and suggestions. This research is financially supported by Vietnam National University, Hanoi - Grant number QG. 23.06.
\section*{References}
\bibliographystyle{iopart-num}
\bibliography{main}

\newpage

\end{document}